\documentclass[aps,prl,showpacs,twocolumn,reprint,amsmath,amssymb,superscriptaddress,floatfix]{revtex4-1}

\usepackage{graphicx}
\usepackage{epstopdf}
\usepackage{amssymb}
\usepackage{amsmath}
\usepackage{bm}
\usepackage{braket}
\usepackage{color}
\usepackage[colorlinks,citecolor=blue]{hyperref}

\begin{document}

\title{Observation of interaction-induced mobility edge in a disordered atomic wire}

\author{Yunfei Wang }
\affiliation{State Key Laboratory of Quantum Optics and Quantum Optics Devices, Institute of Laser Spectroscopy, Shanxi University, Taiyuan 030006, China}

\author{Jia-Hui Zhang}
\affiliation{State Key Laboratory of Quantum Optics and Quantum Optics Devices, Institute of Laser Spectroscopy, Shanxi University, Taiyuan 030006, China}

\author{Yuqing Li }
\thanks{lyqing.2006@163.com}
\affiliation{State Key Laboratory of Quantum Optics and Quantum Optics Devices, Institute of Laser Spectroscopy, Shanxi University, Taiyuan 030006, China}
\affiliation{Collaborative Innovation Center of Extreme Optics, Shanxi University, Taiyuan, Shanxi 030006, China}

\author{Jizhou Wu }
\affiliation{State Key Laboratory of Quantum Optics and Quantum Optics Devices, Institute of Laser Spectroscopy, Shanxi University, Taiyuan 030006, China}
\affiliation{Collaborative Innovation Center of Extreme Optics, Shanxi University, Taiyuan, Shanxi 030006, China}

\author{Wenliang Liu }
\affiliation{State Key Laboratory of Quantum Optics and Quantum Optics Devices, Institute of Laser Spectroscopy, Shanxi University, Taiyuan 030006, China}
\affiliation{Collaborative Innovation Center of Extreme Optics, Shanxi University, Taiyuan, Shanxi 030006, China}

\author{Feng Mei }
\thanks{meifeng@sxu.edu.cn}
\affiliation{State Key Laboratory of Quantum Optics and Quantum Optics Devices, Institute of Laser Spectroscopy, Shanxi University, Taiyuan 030006, China}
\affiliation{Collaborative Innovation Center of Extreme Optics, Shanxi University, Taiyuan, Shanxi 030006, China}

\author{Ying Hu }
\thanks{huying@sxu.edu.cn}
\affiliation{State Key Laboratory of Quantum Optics and Quantum Optics Devices, Institute of Laser Spectroscopy, Shanxi University, Taiyuan 030006, China}
\affiliation{Collaborative Innovation Center of Extreme Optics, Shanxi University, Taiyuan, Shanxi 030006, China}

\author{Liantuan Xiao }
\affiliation{State Key Laboratory of Quantum Optics and Quantum Optics Devices, Institute of Laser Spectroscopy, Shanxi University, Taiyuan 030006, China}
\affiliation{Collaborative Innovation Center of Extreme Optics, Shanxi University, Taiyuan, Shanxi 030006, China}

\author{Jie Ma }
\thanks{mj@sxu.edu.cn}
\affiliation{State Key Laboratory of Quantum Optics and Quantum Optics Devices, Institute of Laser Spectroscopy, Shanxi University, Taiyuan 030006, China}
\affiliation{Collaborative Innovation Center of Extreme Optics, Shanxi University, Taiyuan, Shanxi 030006, China}

\author{Cheng Chin }
\affiliation{James Franck institute, Enrico Fermi institute, Department of Physics, University of Chicago, Illinois 60637, USA}

\author{Suotang Jia  }
\affiliation{State Key Laboratory of Quantum Optics and Quantum Optics Devices, Institute of Laser Spectroscopy, Shanxi University, Taiyuan 030006, China}
\affiliation{Collaborative Innovation Center of Extreme Optics, Shanxi University, Taiyuan, Shanxi 030006, China}

\begin{abstract}
Mobility edge, a critical energy separating localized and extended excitations, is a key concept for understanding quantum localization. Aubry-Andr\'{e} (AA) model, a paradigm for exploring quantum localization, does not naturally allow mobility edges due to self-duality. Using the momentum-state lattice of quantum gas of Cs atoms to synthesize a nonlinear AA model, we provide experimental evidence for mobility edge induced by interactions. By identifying the extended-to-localized transition of different energy eigenstates, we construct a mobility-edge phase diagram. The location of mobility edge in the low- or high-energy region is tunable via repulsive or attractive interactions. Our observation is in good agreement with the theory, and supports an interpretation of such interaction-induced mobility edge via a generalized AA model. Our work also offers new possibilities to engineer quantum transport and phase transitions in disordered systems.
\end{abstract}

\maketitle

\textit{Introduction} -- The concept of mobility edge (ME), a critical energy separating extended and exponentially localized energy eigen-states in the excitation spectrum, is key for understanding Anderson localization~\cite{Anderson1958,Dalichaouch1991,Wiersma1997,Chabanov2000,Strzer2006,Schwartz2007,Julien2008,Lahini2008,Billy2008,Kondov2011,Jendrzejewski2012,Isam2015,White2020} induced by random disorder in three dimensions (3D)~\cite{Abrahams1979}. In low dimensions, arbitrarily weak random disorder can make all single-particle eigenstates localize, hence ME is absent. Localization phenomena have also been actively studied in quasi-periodic lattice systems with incommensurate modulations~\cite{Aubry1980,Grempel1982,Kohmoto1983a,Kohmoto1983b,DasSarma1986,Thouless1988,Modugno2009,
Biddle2009,Biddle2011,Yan2021}, such as those described by Aubry-Andr\'{e} (AA) model~\cite{Aubry1980}.

Realization of the AA model in cold atoms has led to the first observation of the localization transition of a noninteracting Bose-Einstein condensate (BEC)~\cite{Roati2008}. As the AA model has self-dual symmetry~\cite{Aubry1980}, the localization transition is energy-independent (i.e., no ME), with all eigenstates being localized across a single critical point. Intriguingly, variants of AA model which have broken self-duality, such as the  generalized Aubry-Andr\'{e} (GAA) model~\cite{Ganeshan2015}, can host ME already in one dimension (1D). So far, the existence of ME has been mainly conjectured in noninteracting quasi-periodic lattice systems~\cite{DasSarma1988,Goedeke2007,Biddle2010,Ganeshan2015,Li2017,Yao2019,Deng2019,Li2020,Wang2020,Roy2021}, and experimentally confirmed with cold atoms in optical lattices~\cite{Lschen2018}.

\begin{figure}
\includegraphics[width=1\columnwidth]{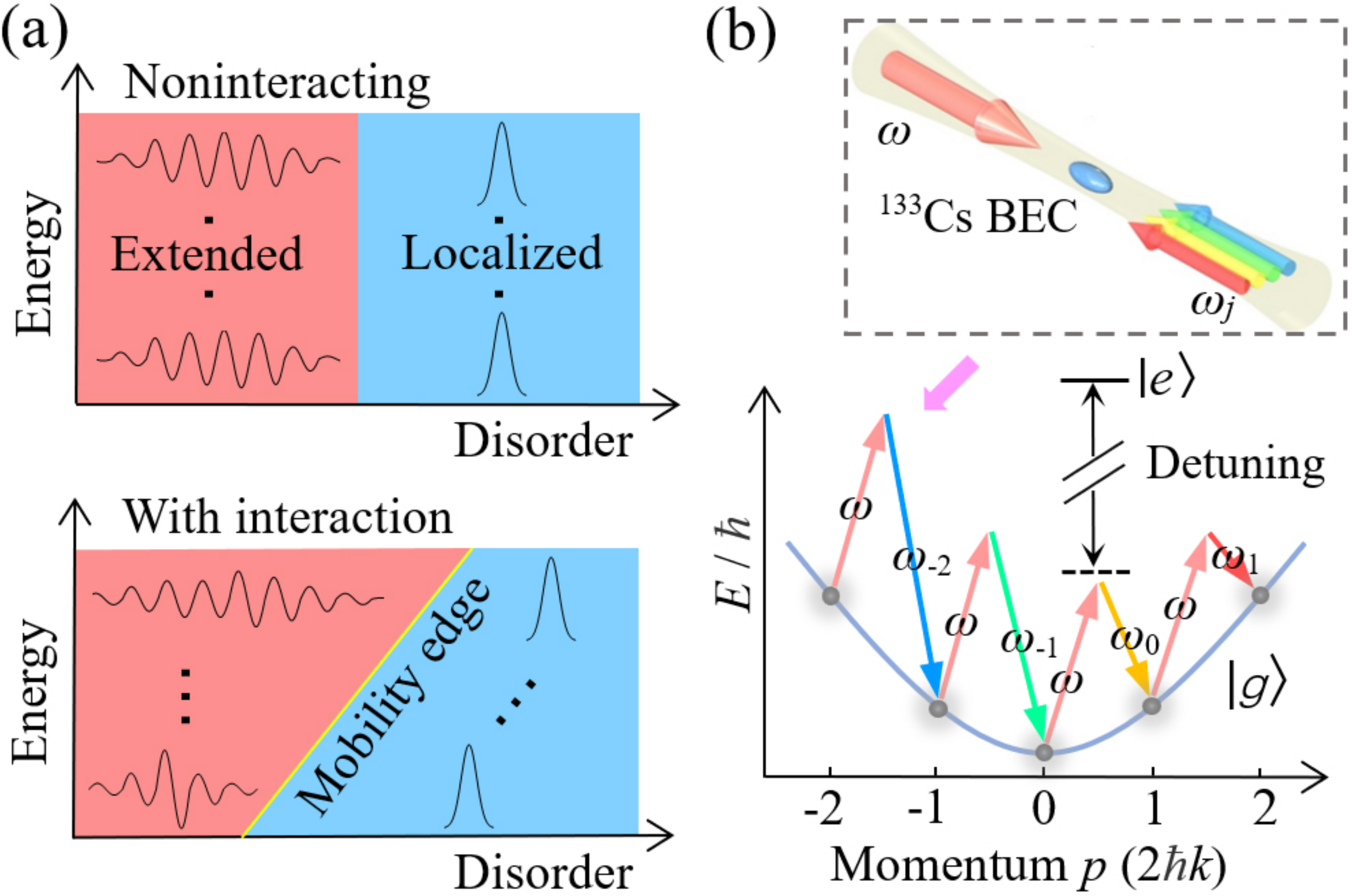}
\caption{\label{Fig1} Illustration of interaction-induced mobility edge and experimental scheme. (a) Top: without interaction, all eigenstates in the energy spectrum are either localized (blue) or extended (red). Bottom: weak interaction can suppress or enhance localization depending on the energy, thus mobility edge emerges in the spectrum. (b) Top: a quasi-1D $^{133}$Cs BEC with tunable atomic interaction is illuminated by a pair of counter-propagating laser beams, one with a frequency $\omega$ and the other with multifrequency components $\omega_j$ ($j=-10,...,9$). Bottom: The lasers, far detuned from the atomic transition, drive a series of engineered two-photon Bragg transitions that couple $21$ momentum states with the increment of $2\hbar k$ (with $k=2\pi/\lambda$). This synthesizes a nonlinear AA model with $L=21$ sites in the momentum space. }
\end{figure}

Beyond noninteracting systems, the realization and control of ME are of fundamental interests, but are generally challenging. Recently, some atomic experiments in this direction have been carried out, showing how single-particle MEs are affected by weak interactions~\cite{An2018a,An2021a}, and many-body ME has been discussed in the context of many-body localization~\cite{Kohlert2019}. In these experiments, however, ME is already expected without interactions. It is thus highly desired to understand MEs based on systems with tunable interactions.

In this work, we demonstrate that ME can be \textit{induced} and \textit{tuned} by interaction; the physical picture is shown in Fig.~\ref{Fig1}(a): weak interaction can have different dressing effects on different energy eigenstates of the AA model, resulting in a suppressed or enhanced localization of an eigenstate, thus a critical energy (i.e., ME) is expected in the excitation spectrum.

Experimentally, we observe signatures of ME based on the momentum-state lattice of quasi-1D $^{133}$Cs BEC that simulates a nonlinear AA model [Fig.~\ref{Fig1}(b)]. Exploiting the tunable scattering length of Cs atoms with Feshbach resonances~\cite{Weber2003,Kraemer2004,Chin2004,Chin2010}, we realize a disordered wire with a wide range of interaction, and observe the extended-to-localized transition in the excitation spectrum. In particular, the controllability of all the system parameters including interactions allows us to access the highest excited state, which can be viewed as the ground state of the associated \textit{negative} Hamiltonian. We demonstrate that interactions can enhance the localization of either low- or high-energy eigenstates, depending on the sign of the interactions. We further construct a mobility-edge phase diagram, which agrees well with the theory.

Such interaction-induced ME can be understood through an effective noninteracting GAA model. While the nonlinear AA model and its variants~\cite{Lahini2009,Deissler2010,Lucioni2011,An2017} have been studied before in optical and atomic setups~\cite{McKenna1992,Pikovsky2008,Deissler2010,Lucioni2011,Lellouch2014,Schwartz2007,Lahini2008,Lahini2009,An2017,Deissler2010,Lucioni2011}, the exploration of ME in the model remains elusive.

\textit{ME in the nonlinear AA model} -- We start by theoretically showing how ME arises in the nonlinear AA model
\begin{eqnarray}
\label{eq:nAA}
i\hbar\dot{\varphi}_{j}&=&J(\varphi _{j+1}+\varphi _{j-1})\nonumber\\
&+&\Delta \cos(2\pi\beta j+\phi)\varphi _{j}-U|\varphi _{j}|^{2}\varphi _{j}.
\end{eqnarray}
Here $\hbar$ is the reduced Planck's constant, ${\varphi}_{j}$ is the wave function at site $j$ in a lattice of size $L$ with $\sum_j |\varphi_j|^2=1$, and $J$ is the nearest-neighbor coupling. The on-site modulation with $\beta=(\sqrt{5}-1)/2$ has an amplitude $\Delta$ and phase $\phi$, which plays the role of disorder. The nonlinear term characterized by $U$ in our subsequent discussion arises from the atomic interaction. When $U=0$, Eq.~(\ref{eq:nAA}) reduces to the AA model, with all eigenstates extended for $\Delta/J<2$ and localized for $\Delta/J>2$.

\begin{figure}
\includegraphics[width=1\columnwidth]{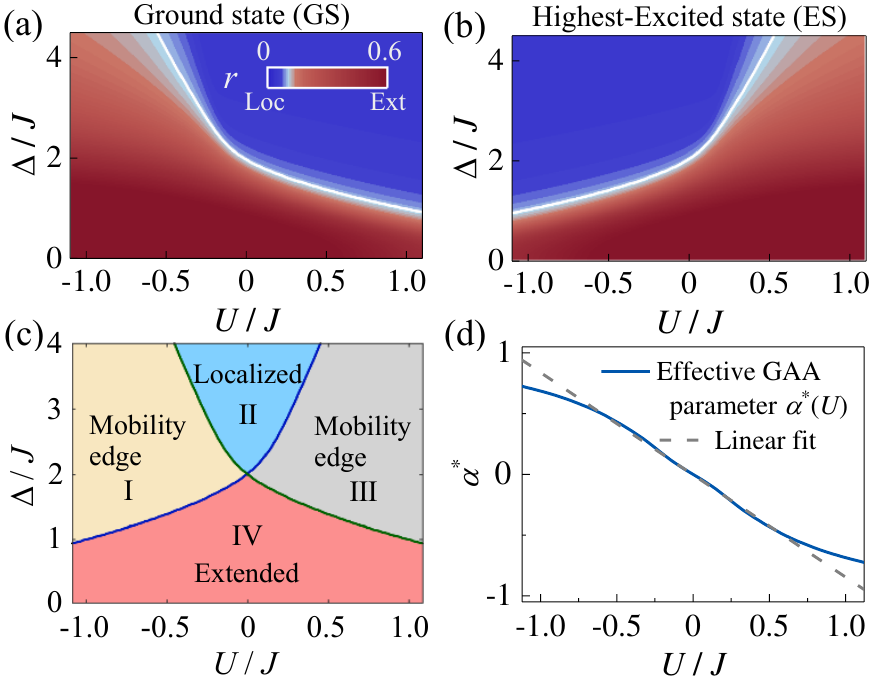}
\caption{\label{Fig2} {Theoretical prediction}. (a)-(b) Participation ratio $r$ of (a) GS and (b) ES for various disorder strength $\Delta/J$ and interaction $U/J$. The state is localized (extended) in blue (red) region. The transition (white curve) is identified as where $r=0.103$, the critical value in the noninteracting case~\cite{footnote2}. (c) Regimes in the parameter spaces $(\Delta/J, U/J)$ where ME may exist. In phases I and III, extended and localized eigenstates coexist, signaling ME; In phase I (III), low energy states are extended (localized). The boundaries are denoted by the green and blue curves. (d) Verification of the effective GAA model. The $\alpha^*$ defined in Eq.~(\ref{eq:Ec}) is calculated as a function of $U/J$ (solid curve). For small $U/J$, it exhibits a linear relation $\alpha^*=-0.81U/J$ (dashed curve). In all panels, the lattice size is $L=21$. }
\end{figure}

\begin{figure*}
\includegraphics[width=0.78\textwidth]{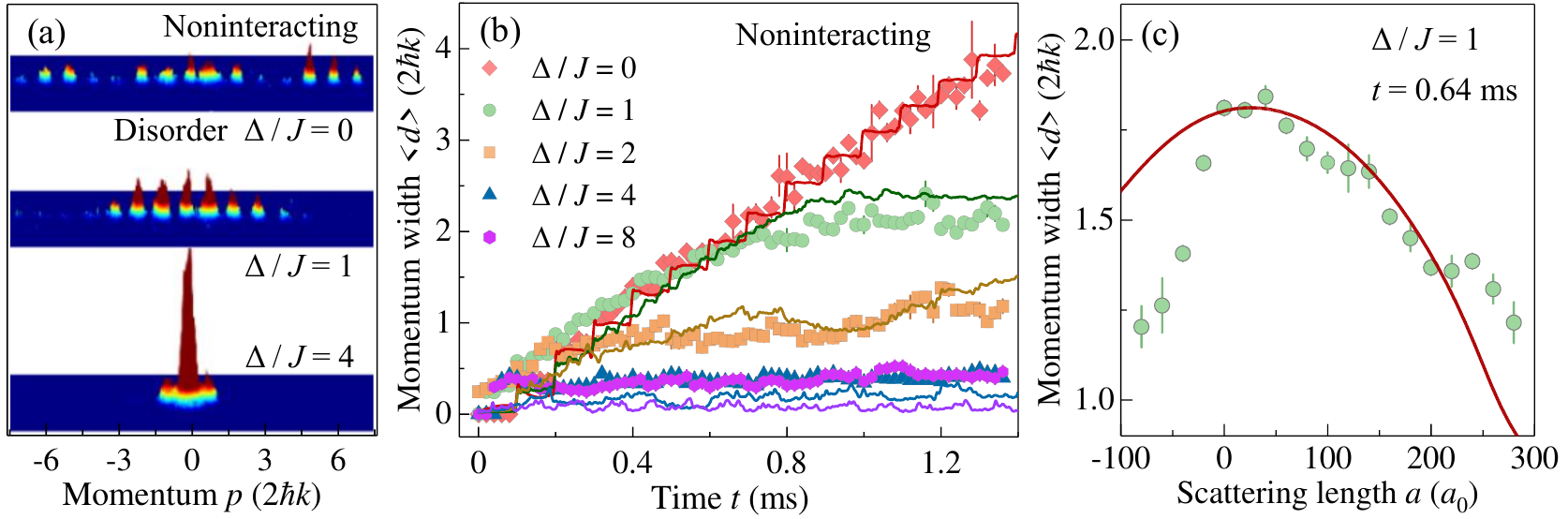}
\caption{\label{Fig3} {Experimental realization of the nonlinear AA model in a Cs BEC with tunable disorder and interaction}. (a) Images of momentum distribution for various disorder strength $\Delta/J$ in a noninteracting BEC. The $\Delta/J$ is tuned via engineering the frequency difference between the pair of Bragg lasers in Fig.~\ref{Fig1}(b). The BEC is initialized at the momentum lattice site $j=0$. The absorption image is taken after $t=1.28$ ms evolution and 18 ms TOF. (b) Measured momentum width  $\langle d(t) \rangle$ as a function of the evolution time $t$ in a noninteracting BEC under various $\Delta/J$. Corresponding solid curves denote the numerical simulations~\cite{Supp}. (c) Measured momentum width  $\langle d \rangle$ at $t= 0.64$ ms as a function of the scattering length $a$ (in unit of Bohr radius $a_0$) for $\Delta/J=1$. The scattering length is tuned via a Feshbach resonance. The solid curve denotes the numerical simulation~\cite{Supp}. Error bars denote $1\sigma$ standard deviations. In all panels, the coupling rates are $J/\hbar=2\pi\times 500$ Hz.}
\end{figure*}

We are interested in the weak interaction regime $|U/J|\lesssim1$ where the self-trapping~\cite{Smerzi1997,Albiez2005} does not occur, and every eigenstate has a correspondence in the noninteracting counterpart~\cite{Wu2005}. Insights can be obtained by noting that the combination of the nonlinear term and the disorder results in a density-dependent potential $V_j^{\textrm{ext}}=\Delta \cos(2\pi\beta j+\phi)-Un_j$, with the density $n_j=|\varphi _{j}|^{2}$. As the density distribution is shaped by the incommensurate modulation, $V_j^{\textrm{ext}}$ contains \textit{multiple} harmonics of the quasi-periodicity (i.e., $2\pi\beta$), which breaks the self-duality and leads to ME. For $|U/\Delta|\ll 1$, perturbative analysis suggests $V_j^{\textrm{ext}}$ is effectively a GAA lattice potential. For instance, when $\phi=0$, by Fourier expanding the density up to the second harmonics of the quasi-periodicity, we have $V_j^{\textrm{ext}}\approx(\Delta-Uc_1)\cos(2\pi\beta j)-Uc_2\cos(4\pi\beta j)$ (apart from some constant), with the expansion coefficients $c_{1}$ and $c_2$. It approximates the GAA lattice potential~\cite{Ganeshan2015}
\begin{equation}
V^j_\textrm{GAA}=\Delta\frac{\cos(2\pi\beta j)}{1-\alpha^*(U)\cos(2\pi\beta j)}\label{eq:VGAA}
\end{equation}
with $\alpha^*\ll 1$ and $\alpha^*\propto -U$, up to the second harmonics. Thus the physics of Eq.~(\ref{eq:nAA}) may be understood via an \textit{effective noninteracting GAA model}: $
i\hbar\dot{\varphi}_{j}=\!\!J(\varphi _{j+1}+\varphi _{j-1})\!+\!V^j_\textrm{GAA}\varphi _{j}$.
As the GAA model hosts an exact ME~\cite{Ganeshan2015}, the location $E_c$ of ME in a weakly nonlinear AA model is expected to be
\begin{equation}
E_c=\frac{\textrm{sgn}(\Delta)(2|J|-|\Delta|)}{\alpha^*(U)},\label{eq:Ec}
\end{equation}
where $\textrm{sgn}$ denotes the sign function. Because $\alpha^*\propto -U$, we expect the location of ME in the low- or high-energy region is swapped when $U\rightarrow -U$.

The above analysis is supported by numerical calculations. We focus on the ground state (GS) and the highest excited state (ES) of the nonlinear AA model~\cite{footnote1}, and characterize their degree of localization via the participation ratio
\begin{equation}
r=\frac{1}{L}\frac{1}{\sum _{j=1}^L n_{j}^{2}}. \label{eq:PR}
\end{equation}
For a localized state, $r\approx 0$; for an extended state, the maximum possible $r$ is $1$. Figure~\ref{Fig2}(a) and (b) show the participation ratio $r$ as a function of $\Delta/J$ and $U/J$ for GS and ES, respectively. To identify the transition from the extended to the localized, we use the critical value of $r$ at $\Delta/J=2$ in the noninteracting limit (see white curves)~\cite{footnote2}; for $L=21$ sites, the critical value is $r=0.103$. We see that while the transition points of GS and ES coincide without interaction, they differ in the presence of interaction, suggesting the transition is energy-dependent. Moreover, adding $U>0$ enhances localization of GS but delocalizes ES, whereas the opposite occurs for $U<0$. The critical points of GS and ES divide the parameter space $(\Delta, U)$ into four regimes [Fig.~\ref{Fig2}(c)]. In phase II (IV), all states are localized (extended). But in both phases I and III, extended and localized states coexist, signaling the existence of ME. When increasing disorder, ME emerges from the high-energy regime in I ($U<0$), while it emerges from the low-energy regime in III ($U>0$).

We validate the effective GAA model through numerical simulations. We calculate the effective parameter $\alpha^*=(\Delta_c^g-\Delta_c^e)/(E^c_e-E^c_g)$ based on Eq.~(\ref{eq:Ec}), where $E^c_{g}$ ($E^c_{e}$) denotes the energy of GS (ES) at their critical point $\Delta_c^{g}$ ($\Delta_c^{e}$), representing where ME coincides with the lowest (highest) energy. Figure~\ref{Fig2}(d) shows $\alpha^*$ as a function of $U/J$ (blue curve). It is linear for small interactions, confirming the previous conjecture. In the linear regime, we expect the location of ME to be given by Eq.~(\ref{eq:Ec}). When the linearity breaks down, the effective model is no longer suitable. \\

\textit{Experimental realization of the model} -- We experimentally realize the nonlinear AA model using the momentum-state lattice of $^{133}$Cs BEC that contains $4\times 10^4$ atoms in the hyperfine state $|F = 3, m_{F} = 3 \rangle$ [Fig.~\ref{Fig1}(b)]~\cite{Wang2021}. We start with a BEC confined in a quasi-1D optical trap~\cite{Supp}. Two counter-propagating laser beams with the wavelength $\lambda = 1064$ nm are applied, one with a frequency $\omega$, while the other containing multi-frequency components $\omega_j =\omega-\Delta\omega_j$, $j=-10,...,9$. They drive a series of two-photon Bragg transitions to couple $21$ discrete momentum states with the momentum increment of $2\hbar k$ (with $k=2\pi/\lambda$), which simulates AA model of $L=21$ sites with the nearest-neighbor coupling $J$~\cite{Meier2016}. The disorder is realized by engineering $\Delta \omega_j$ to yield an on-site energy $\Delta \cos(2\beta\pi j + \phi)$~\cite{Supp} with both $\Delta$ and $\phi$ controllable via Bragg lasers. We employ a broad Feshbach resonance centered at magnetic field $B=-11.7$ G to tune the atomic $s$-wave scattering length~\cite{Weber2003,Kraemer2004,Chin2004,Chin2010}. According to the mean-field theory of the momentum-lattice system~\cite{Chen,An2021a,An2018,An2021b}, the atomic interaction leads to the nonlinear term in Eq.~(\ref{eq:nAA}), with $U=(4\pi\hbar^2a/m)\bar{\rho}$, where $m$ is the atomic mass and $\bar{\rho}$ is an effective atomic density~\cite{Supp}. We note that a quasi-1D BEC can be stable for $a<0$~\cite{BECBOOK}; Experimentally, we do not observe the collapse of BEC for $a<0$ on the time scale of $2$ ms relevant for our measurements. To avoid significant three-body loss~\cite{Kraemer2006}, we restrict ourselves to $a>-100a_0$ ($a_0$ is the Bohr radius).

\begin{figure}
\includegraphics[width=0.95\columnwidth]{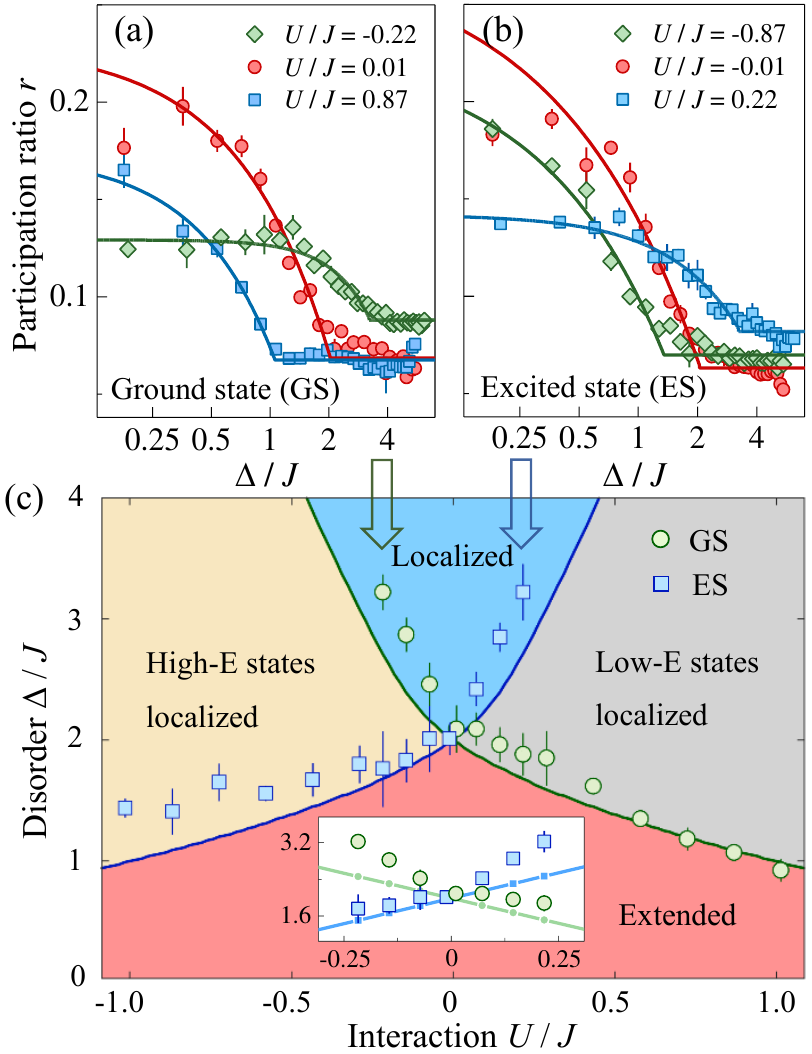}
\caption{\label{Fig4} {Construction of mobility-edge (ME) phase diagram}. (a)-(b) Identification of the extended-to-localized transition points of (a) GS and (b) ES. The experimental participation ratio $r$ is shown as a function of disorder $\Delta/J$ in the semi-$\log_2$ plot for various interactions. By fitting (solid curves) the experimental data via an empirical relation~\cite{Supp}, we extract the transition point. In (a) and (b), the error bar indicates the $1\sigma$ standard deviation. (c) Construction of the phase diagram and identification of ME. We superimpose the extracted transition points of GS (green circles) and ES (blue squares) on the theoretical phase diagram in Fig.~\ref{Fig2}(c). In the inset, we compare the data for $|U/J|<0.25$ with the predictions from the effective GAA model [green (blue) line for GS (ES)]. The error bar in (c) shows the fitting error. In (a)-(c), the coupling rates are $J/\hbar=2\pi\times 275$ Hz.}
\end{figure}

To confirm the realization of the nonlinear AA model, we first tune $a\approx 0$ and observe the transport of the noninteracting BEC by controlling the disorder strength. Initially, the BEC with zero momentum is prepared at momentum-state lattice site $j=0$, before the Bragg lasers are switched on. After an evolution time $t$, the momentum distribution is measured through the time-of-flight (TOF) technique~\cite{Supp}. TOF images for $t=4\hbar/J=1.28$ ms under different disorder strength are illustrated in Fig.~\ref{Fig3}(a). As expected from the AA model, atoms spread over several lattice sites for $\Delta/J<2$, but localizing sharply near the initial position for $\Delta/J>2$. Figure~\ref{Fig3}(b) measures the time-dependent momentum width, $\langle d (t) \rangle=\sum_j |j| n_j(t)$, where $n_j(t)$ is the measured fraction of atomic population at the momentum state $|2j\hbar k\rangle$ at time $t$. We observe the familiar crossover from the ballistic transport to localization with increasing disorder, in agreement with the numerical simulations (solid lines) based on Gross-Pitaevskii (GP) equation~\cite{Supp}.

Next, we tune the scattering length from $a<0$ to $a>0$ using the broad Feshbach resonance and fix the disorder at $\Delta/J=1$. Figure~\ref{Fig3}(c) shows the measured momentum width $\langle d\rangle$ at $t=2\hbar/J=0.64$ ms under various interactions. We observe $\langle d\rangle$ decreases with the interaction strength, regardless of its sign, suggesting an increased degree of localization. The experiment agrees qualitatively with the GP calculations~\cite{Supp} (red curve). \\

\textit{Construction of ME phase diagram} -- Important for our study is the preparation of GS and ES under various disorder and interaction. We prepare GS following the adiabatic protocol in Ref.~\cite{An2021a}. It consists of switching off the laser coupling between lattice sites and initializing atoms in the ground state at zero-momentum site $j=0$. Then, by linearly ramping the coupling from $J=0$ to $J/\hbar= 2\pi\times 275$ Hz in $1$ ms and holding there for $1$ ms, the initial state is transferred to GS of the lattice system.

A reliable preparation of ES, however, was nontrivial due to non-adiabatic effects~\cite{An2021a}. Here we adopt a strategy which is motivated by the fact that ES of a Hamiltonian $H$ can be viewed as GS of $-H$. Experimentally, in preparing ES of a system with desired $J$, $\Delta$ and $a$ we instead realize an associated ``negative Hamiltonian''  (i.e., $-H$) with $-J$, $-\Delta$ and $-a$. To realize $-J$, we introduce a relative phase $\pi$ in the two-photon Bragg transition coupling neighboring momentum states [c.f. Fig.~\ref{Fig1}(b)]. The $-\Delta$ is achieved by tuning $\phi$. Crucially, the conversion from $a$ to $-a$ is uniquely enabled by Feshbach tuning of the interaction of Cs atoms. Then we adiabatically prepare GS of the negative system similarly as before: after switching off the laser coupling and initializing atoms at a site with the lowest energy, we linearly ramp up $J$, thus achieving ES of the original system.

After the state preparation, we measure the population in each momentum mode to obtain the participation ratio $r$ according to Eq.~(\ref{eq:PR}). We identify the potential extended-to-localized transition by numerically fitting the experimental data of $r$ as a function of $\Delta/J$~\cite{Supp}, as shown in Figs.~\ref{Fig4}(a)-(b). For $U\approx 0$, the transition is at $\Delta/J=2$ for both GS and ES, as expected. Comparing Figs.~\ref{Fig4}(a) and (b) shows that the localization of GS is enhanced (suppressed) under $U>0$ ($U<0$), whereas the opposite occurs for ES, in agreement with the predictions. Owing to the non-adiabatic effect in the experimental ramp, the participation ratio $r$ is generically smaller than the idealized value expected from Eq.~(\ref{eq:nAA})~\cite{Supp}. Moreover, the experimental $r$ deep in the localized phase is slightly higher than $B=1/21$ due to the residual population. Nevertheless, these imperfections and finite-size effect do not qualitatively change the transitions~\cite{Supp}.

Finally, to construct the phase diagram, we collect the extracted transition points of GS and ES under various interactions into Fig.~\ref{Fig4}(c). The good agreement between the experiment and theory provides strong evidences on the existence of interaction-induced ME. It also confirms that, depending on whether the atomic interaction $U$ is attractive or repulsive, ME emerges from the high- or low-energy region of the excitation spectrum. In the inset of Fig.~\ref{Fig4}(c), we compare the experimental data with the predictions from the effective GAA model [c.f. Fig.~\ref{Fig2}(d)]. As shown, the effective model provides a good explanation of the data for $|U/J|\lesssim 0.25$, suggesting the location of ME in this regime is given by Eq.~(\ref{eq:Ec}).

\textit{Conclusion} -- In this work we have provided experimental evidence that ME can be induced and controlled by interactions. The interaction-induced ME is different from the single-particle ME in noninteracting models and from the many-body ME~\cite{Abanin2019,Deng2017,Li2015}. Our observations shed new light on the interplay between disorder and interaction. The widely tunable atomic interaction featured in our experiment enables the first observation of this intriguing phenomena, which presents new opportunities for engineering the quantum transport and quantum phase transitions in disordered systems.

\textit{Acknowledgement} -- We acknowledge helpful discussions with Biao Wu, Xiaopeng Li, Zhihao Xu and Shuai Chen. This research is funded by National Key R\&D Program of China (Grant No. 2017YFA0304203), National Natural Science Foundation of China (Grant No. 62020106014, 92165106, 62175140, 12104276, 11874038, 12034012, 12074234). CC acknowledges support by the National Science Foundation (Grant No. PHY-2103542).

~\\


\begin{thebibliography}{10}

\bibitem{Anderson1958} P. W. Anderson, Absence of diffusion in certain random lattices, Phys. Rev. \textbf{109}, 1492 (1958).

\bibitem{Dalichaouch1991} R. Dalichaouch, J. P. Armstrong, S. Schultz, P. M. Platzman, and S. L. McCall, Microwave localization by two-dimensional random scattering, Nature \textbf{354}, 53 (1991).

\bibitem{Wiersma1997} D. Wiersma, P. Bartolini, A. Lagendijk, and R. Righini, Localization of light in a disordered medium, Nature \textbf{390}, 671 (1997).

\bibitem{Chabanov2000} A. A. Chabanov, M. Stoytchev, and A. Z. Genack, Statistical signatures of photon localization, Nature \textbf{404}, 850 (2000).

\bibitem{Strzer2006} M. St\"{o}rzer, P. Gross, C. Aegerter, and G. Maret, Observation of the critical regime near Anderson localization of light, Phys. Rev. Lett. \textbf{96}, 063904 (2006).

\bibitem{Schwartz2007} T. Schwartz, G. Bartal, S. Fishman, and M. Segev, Transport and Anderson localization in disordered two-dimensional photonic lattices, Nature \textbf{446}, 52 (2007).

\bibitem{Julien2008} J. Chab\'{e}, G. Lemari\'{e}, B. Gr\'{e}maud, D. Delande, P. Szriftgiser, and J. C. Garreau, Experimental observation of the Anderson metal-insulator transition with atomic matter waves, Phys. Rev. Lett. \textbf{101}, 255702 (2008).

\bibitem{Lahini2008} Y. Lahini, A. Avidan, F. Pozzi, M. Sorel, R. Morandotti, D. Christodoulides, and Y. Silberberg, Anderson localization and nonlinearity in one-dimensional disordered photonic lattices, Phys. Rev. Lett. \textbf{100}, 013906 (2008).

\bibitem{Billy2008} J. Billy, V. Josse, Z. C. Zuo, A. Bernard, B. Hambrecht, P. Lugan, D. Cl\'{e}ment, L. Sanchez-Palencia, P. Bouyer, and A. Aspect, Direct observation of Anderson localization of matter waves in a controlled disorder, Nature \textbf{453}, 891 (2008).

\bibitem{Kondov2011} S. S. Kondov, W. R. McGehee, J. J. Zirbel, and B. DeMarco, Three-dimensional Anderson localization of ultracold matter, Science \textbf{334}, 66 (2011).

\bibitem{Jendrzejewski2012} F. Jendrzejewski, A. Bernard, K. M\"{u}ller, P. Cheinet, V. Josse, M. Piraud, L. Pezz\'{e}, L. Sanchez-Palencia, A. Aspect, and P. Bouyer, Three-dimensional localization of ultracold atoms in an optical disordered potential, Nat. Phys. \textbf{8}, 398 (2012).

\bibitem{Isam2015} I. Manai, J. Cl\'{e}ment, R. Chicireanu, C. Hainaut, J. C. Garreau, P. Szriftgiser, and D. Delande, Experimental observation of two-dimensional Anderson localization with the atomic kicked rotor, Phys. Rev. Lett. \textbf{115}, 240603 (2015).


\bibitem{White2020} D. White, T. Haase, D. Brown, M. Hoogerland, M. Najafabadi, J. Helm, C. Gies, D. Schumayer, and D. W. Hutchinson, Observation of two-dimensional Anderson localisation of ultracold atoms, Nat. Commun. \textbf{11}, 4942 (2020).

\bibitem{Abrahams1979} E. Abrahams, P. W. Anderson, D. C. Licciardello, and T. V. Ramakrishnan, Scaling theory of localization: Absence of quantum diffusion in two dimensions, Phys. Rev. Lett. \textbf{42}, 673 (1979).

\bibitem{Aubry1980} S. Aubry and G. Andr\'{e}, Analyticity breaking and Anderson localization in incommensurate lattices, Ann. Israel Phys. Soc. \textbf{3}, 18 (1980).

\bibitem{Grempel1982} D. R. Grempel, S. Fishman, and R. E. Prange, Localization in an incommensurate potential: An exactly solvable model, Phys. Rev. Lett. \textbf{49}, 833 (1982).

\bibitem{Kohmoto1983a} M. Kohmoto, L. Kadanoff, and C. Tang, Localization problem in one dimension: Mapping and escape, Phys. Rev. Lett. \textbf{50}, 1870 (1983).

\bibitem{Kohmoto1983b} M. Kohmoto, Metal-insulator transition and scaling for incommensurate systems, Phys. Rev. Lett. \textbf{51}, 1198 (1983).

\bibitem{DasSarma1986} S. D. Sarma, A. Kobayashi, and R. E. Prange, Proposed experimental realization of Anderson localization in random and incommensurate artificially layered systems, Phys. Rev. Lett. \textbf{56}, 1280 (1986).

\bibitem{Thouless1988} D. J. Thouless, Localization by a potential with slowly varying period, Phys. Rev. Lett. \textbf{61}, 2141 (1988).

\bibitem{Modugno2009} M. Modugno, Exponential localization in one-dimensional quasi-periodic optical lattices, New J. Phys. \textbf{11}, 033023 (2009).


\bibitem{Biddle2009} J. Biddle, B. Wang, D. J. Priour, Jr., and S. D. Sarma, Localization in one-dimensional incommensurate lattices beyond the Aubry-Andr\'{e} model, Phys. Rev. A \textbf{80}, 021603(R) (2009).

\bibitem{Biddle2011} J. Biddle, D. J. Priour, Jr., B. Wang, and S. D. Sarma, Localization in one-dimensional lattices with non-nearest-neighbor hopping: Generalized Anderson and Aubry-Andr\'{e} models, Phys. Rev. B \textbf{83}, 075105 (2011).

\bibitem{Yan2021} T. Xiao, D. Z. Xie, Z. L. Dong, T. Chen, W. Yi, and B. Yan, Observation of topological phase with critical localization in a quasi-periodic lattice, Sci. Bull. \textbf{66}, 2175 (2021).


\bibitem{Roati2008} G. Roati, C. D'Errico, L. Fallani, M. Fattori, C. Fort, M. Zaccanti, G. Modugno, M. Modugno, and M. Inguscio, Anderson localization of a non-interacting Bose-Einstein condensate, Nature \textbf{453}, 895 (2008).

\bibitem{Ganeshan2015} S. Ganeshan, J. H. Pixley, and S. D. Sarma, Nearest neighbor tight binding models with an exact mobility edge in one dimension, Phys. Rev. Lett. \textbf{114}, 146601 (2015).

\bibitem{DasSarma1988} S. D. Sarma, S. He, and X. C. Xie, Mobility edge in a model one-dimensional potential, Phys. Rev. Lett. \textbf{61}, 2144 (1988).

\bibitem{Goedeke2007} D. J. Boers, B. Goedeke, D. Hinrichs, and M. Holthaus, Mobility edges in bichromatic optical lattices, Phys. Rev. A \textbf{75}, 063404 (2007).

\bibitem{Biddle2010} J. Biddle and S. D. Sarma, Predicted mobility edges in one-dimensional incommensurate optical lattices: An exactly solvable model of Anderson localization, Phys. Rev. Lett. \textbf{104}, 070601 (2010).

\bibitem{Li2017} X. Li, X. P. Li, and S. D. Sarma, Mobility edges in one-dimensional bichromatic incommensurate potentials, Phys. Rev. B \textbf{96}, 085119 (2017).

\bibitem{Yao2019} H. P. Yao, H. Khoudli, L. Bresque, and L. Sanchez-Palencia, Critical behavior and fractality in shallow one-dimensional quasiperiodic potentials, Phys. Rev. Lett. \textbf{123}, 070405 (2019).

\bibitem{Deng2019} X. Deng, S. Ray, S. Sinha, G. V. Shlyapnikov, and L. Santos, One-dimensional quasicrystals with Power-Law hopping, Phys. Rev. Lett. \textbf{123}, 025301 (2019).

\bibitem{Li2020} X. Li and S. D. Sarma, Mobility edge and intermediate phase in one-dimensional incommensurate lattice potentials, Phys. Rev. B \textbf{101}, 064203 (2020).

\bibitem{Wang2020} Y. C. Wang, X. Xia, L. Zhang, H. P. Yao, S. Chen, J. G. You, Q. Zhou, and X. J. Liu, One-dimensional quasiperiodic mosaic lattice with exact mobility edges, Phys. Rev. Lett. \textbf{125}, 196604 (2020).

\bibitem{Roy2021} S. Roy, T. Mishra, B. Tanatar, and S. Basu, Reentrant localization transition in a quasiperiodic chain, Phys. Rev. Lett. \textbf{126}, 106803 (2021).

\bibitem{Lschen2018} H. L\"{u}schen, S. Scherg, T. Kohlert, M. Schreiber, P. Bordia, X. Li, S. D. Sarma, and I. Bloch, Single-particle mobility edge in a one-dimensional quasiperiodic optical lattice, Phys. Rev. Lett. \textbf{120}, 160404 (2018).

\bibitem{An2018a} F. A. An, E. J. Meier, and B. Gadway, Engineering a flux-dependent mobility edge in disordered Zigzag chains, Phys. Rev. X \textbf{8}, 031045 (2018).

\bibitem{An2021a} F. A. An, K. Padavi\'{c}, E. J. Meier, S. Hegde, S. Ganeshan, J. H. Pixley, S. Vishveshwara, and B. Gadway, Interactions and mobility edges: Observing the generalized Aubry-Andr\'{e} model, Phys. Rev. Lett. \textbf{126}, 040603 (2021).

\bibitem{Kohlert2019} T. Kohlert, S. Scherg, X. Li, H.  L\"{u}schen, S. D. Sarma, I. Bloch, and M. Aidelsburger, Observation of many-body localization in a one-dimensional system with a single-particle mobility edge, Phys. Rev. Lett. \textbf{122}, 170403 (2019).

\bibitem{Weber2003} T. Weber, J. Herbig, M. Mark, H. -C. N\"{a}gerl, and R. Grimm, Bose-Einstein condensation of cesium, Science \textbf{299}, 232 (2003).

\bibitem{Kraemer2004} T. Kraemer, J. Herbig, M. Mark, T. Weber, C. Chin, H. -C. N\"{a}gerl, and R. Grimm, Optimized production of a cesium Bose-Einstein condensate, Appl. Phys. B \textbf{79}, 1013 (2004).

\bibitem{Chin2004} C. Chin, V. Vuleti\'{c}, A. Kerman, S. Chu, E. Tiesinga, P. Leo, and C. Williams, Precision feshbach spectroscopy of ultracold Cs$_{2}$, Phys. Rev. A \textbf{70}, 032701 (2004).

\bibitem{Chin2010} C. Chin, R. Grimm, P. Julienne, and E. Tiesinga, Feshbach resonances in ultracold gases, Rev. Mod. Phys. \textbf{82}, 1225 (2010).

\bibitem{Lahini2009} Y. Lahini, R. Pugatch, F. Pozzi, M. Sorel, R. Morandotti, N. Davidson, and Y. Silberberg, Observation of a localization transition in quasiperiodic photonic lattices, Phys. Rev. Lett. \textbf{103}, 013901 (2009).

\bibitem{An2017} F. A. An, E. J. Meier, and B. Gadway, Diffusive and arrested transport of atoms under tailored disorder, Nat. Commun. \textbf{8}, 325 (2017).

\bibitem{Deissler2010} B. Deissler, M. Zaccanti, G. Roati, C. D'Errico, M. Fattori, M. Modugno, G. Modugno, and M. Inguscio, Delocalization of a disordered bosonic system by repulsive interactions, Nat. Phys. \textbf{6}, 354 (2010).

\bibitem{Lucioni2011} E. Lucioni, B. Deissler, L. Tanzi, G. Roati, M. Zaccanti, M. Modugno, M. Larcher, F. Dalfovo, M. Inguscio, and G. Modugno, Observation of subdiffusion in a disordered interacting system, Phys. Rev. Lett. \textbf{106}, 230403 (2011).

\bibitem{McKenna1992} M. J. McKenna, R. L. Stanley, and J. D. Maynard, Effects of nonlinearity on Anderson localization, Phys. Rev. Lett. \textbf{69}, 1807 (1992).

\bibitem{Pikovsky2008} A. S. Pikovsky and D. L. Shepelyansky, Destruction of Anderson localization by a weak nonlinearity, Phys. Rev. Lett. \textbf{100}, 094101 (2008).

\bibitem{Lellouch2014} S. Lellouch and L. Sanchez-Palencia, Localization transition in weakly interacting Bose superfluids in one-dimensional quasiperdiodic lattices, Phys. Rev. A \textbf{90}, 061602(R) (2014).

\bibitem{Smerzi1997} A. Smerzi, S. Fantoni, S. Giovanazzi, and S. R. Shenoy, Quantum coherent atomic tunneling between two trapped Bose-Einstein condensates, Phys. Rev. Lett. \textbf{79}, 4950 (1997).

\bibitem{Albiez2005} M. Albiez, R. Gati, J. F\"{o}lling, S. Hunsmann, M. Cristiani, and M. Oberthaler, Direct observation of tunneling and nonlinear self-trapping in a single bosonic Josephson junction, Phys. Rev. Lett. \textbf{95}, 010402 (2005).

\bibitem{Wu2005} B. Wu and Q. Niu, Superfluidity of Bose-Einstein condensate in an optical lattice: Landau-Zener tunnelling and dynamical instability, New. J. Phys. \textbf{5}, 104 (2003).

\bibitem{footnote1} To numerically calculate the highest excited state, we invert the sign of all the system parameters in the nonlinear AA model and use imaginary time evolution to find its ground state.

\bibitem{footnote2} This provides a convenient way for numerically identifying the critical point when the system is small. It is consistent with the more generic approach using power-law fitting to identify the transition, as we did later in the experiment.

\bibitem{Wang2021} Y. F. Wang, Y. Q. Li, J. Z. Wu, W. L. Liu, J. Z. Hu, J. Ma, L. T. Xiao, and S. T. Jia, Hybrid evaporative cooling of $^{133}$Cs atoms to Bose-Einstein condensation, Opt. Express \textbf{29}, 13960 (2021).

\bibitem{Supp} Details are provided in the Supplementary Materials.

\bibitem{Meier2016} E. J. Meier, F. A. An, and B. Gadway, Atom-optics simulator of lattice transport phenomena, Phys. Rev. A \textbf{93}, 051602(R) (2016).

\bibitem{Chen} T. Chen, D. Z. Xie, B. Gadway, and B. Yan, A Gross-Pitaevskii-equation description of the momentum-state lattice: roles of the trap and many-body interactions. arXiv:2103.14205v2.

\bibitem{An2018} F. A. An, E. J. Meier, J. Ang'ong'a, and B. Gadway, Correlated dynamics in a synthetic lattice of momentum states, Phys. Rev. Lett. \textbf{120}, 040407 (2018).

\bibitem{An2021b} F. A. An, B. Sundar, J. P. Hou, X. W. Luo, E. J. Meier, C. W. Zhang, K. A. Hazzard, and B. Gadway, Nonlinear dynamics in a synthetic momentum-state lattice, Phys. Rev. Lett. \textbf{127}, 130401 (2021).


\bibitem{BECBOOK} C. J. Pethick and H. Smith, Bose-Einstein Condensation in Dilute Gases (Cambridge University Press, Cambridge, 2008).

\bibitem{Kraemer2006} T. Kraemer, M. Mark, P. Waldburger, J. G. Danzl, C. Chin, B. Engeser, A. D. Lange, K. Pilch, A. Jaakkola, H. -C. N\"{a}gerl, and R. Grimm, Evidence for Efimov quantum states in an ultracold gas of caesium atoms, Nature \textbf{440}, 315 (2006).

\bibitem{Abanin2019} D. A. Abanin, E. Altman, I. Bloch, and M. Serbyn, Many-body localization, thermalization, and entanglement, Rev. Mod. Phys. \textbf{91}, 021001 (2019).

\bibitem{Deng2017} D. L. Deng, S. Ganeshan, X. P. Li, R. Modak, S. Mukerjee, and J. H. Pixley, Many-body localization in incommensurate models with a mobility edge, Ann. Phys. \textbf{529}, 1600399 (2017).

\bibitem{Li2015} X. P. Li, S. Ganeshan, J. H. Pixley, and S. D. Sarma, Many-body localization and quantum nonergodicity in a model with a single-particle mobility edge, Phys. Rev. Lett. \textbf{115}, 186601 (2015).


\end{thebibliography}
\end{document}


\title{Supplementary Materials for ``Observation of interaction-induced mobility edge in a disordered atomic wire''}

\author{Yunfei Wang }
\affiliation{State Key Laboratory of Quantum Optics and Quantum Optics Devices, Institute of Laser Spectroscopy, Shanxi University, Taiyuan 030006, China}

\author{Jia-Hui Zhang}
\affiliation{State Key Laboratory of Quantum Optics and Quantum Optics Devices, Institute of Laser Spectroscopy, Shanxi University, Taiyuan 030006, China}

\author{Yuqing Li }
\thanks{lyqing.2006@163.com}
\affiliation{State Key Laboratory of Quantum Optics and Quantum Optics Devices, Institute of Laser Spectroscopy, Shanxi University, Taiyuan 030006, China}
\affiliation{Collaborative Innovation Center of Extreme Optics, Shanxi University, Taiyuan, Shanxi 030006, China}

\author{Jizhou Wu }
\affiliation{State Key Laboratory of Quantum Optics and Quantum Optics Devices, Institute of Laser Spectroscopy, Shanxi University, Taiyuan 030006, China}
\affiliation{Collaborative Innovation Center of Extreme Optics, Shanxi University, Taiyuan, Shanxi 030006, China}

\author{Wenliang Liu }
\affiliation{State Key Laboratory of Quantum Optics and Quantum Optics Devices, Institute of Laser Spectroscopy, Shanxi University, Taiyuan 030006, China}
\affiliation{Collaborative Innovation Center of Extreme Optics, Shanxi University, Taiyuan, Shanxi 030006, China}

\author{Feng Mei }
\thanks{meifeng@sxu.edu.cn}
\affiliation{State Key Laboratory of Quantum Optics and Quantum Optics Devices, Institute of Laser Spectroscopy, Shanxi University, Taiyuan 030006, China}
\affiliation{Collaborative Innovation Center of Extreme Optics, Shanxi University, Taiyuan, Shanxi 030006, China}

\author{Ying Hu }
\thanks{huying@sxu.edu.cn}
\affiliation{State Key Laboratory of Quantum Optics and Quantum Optics Devices, Institute of Laser Spectroscopy, Shanxi University, Taiyuan 030006, China}
\affiliation{Collaborative Innovation Center of Extreme Optics, Shanxi University, Taiyuan, Shanxi 030006, China}

\author{Liantuan Xiao }
\affiliation{State Key Laboratory of Quantum Optics and Quantum Optics Devices, Institute of Laser Spectroscopy, Shanxi University, Taiyuan 030006, China}
\affiliation{Collaborative Innovation Center of Extreme Optics, Shanxi University, Taiyuan, Shanxi 030006, China}

\author{Jie Ma }
\thanks{mj@sxu.edu.cn}
\affiliation{State Key Laboratory of Quantum Optics and Quantum Optics Devices, Institute of Laser Spectroscopy, Shanxi University, Taiyuan 030006, China}
\affiliation{Collaborative Innovation Center of Extreme Optics, Shanxi University, Taiyuan, Shanxi 030006, China}

\author{Cheng Chin }
\affiliation{James Franck institute, Enrico Fermi institute, Department of Physics, University of Chicago, Illinois 60637, USA}

\author{Suotang Jia  }
\affiliation{State Key Laboratory of Quantum Optics and Quantum Optics Devices, Institute of Laser Spectroscopy, Shanxi University, Taiyuan 030006, China}
\affiliation{Collaborative Innovation Center of Extreme Optics, Shanxi University, Taiyuan, Shanxi 030006, China}

\maketitle

\subsection{Experimental setup}

\begin{figure}
	\includegraphics[width=0.9\columnwidth]{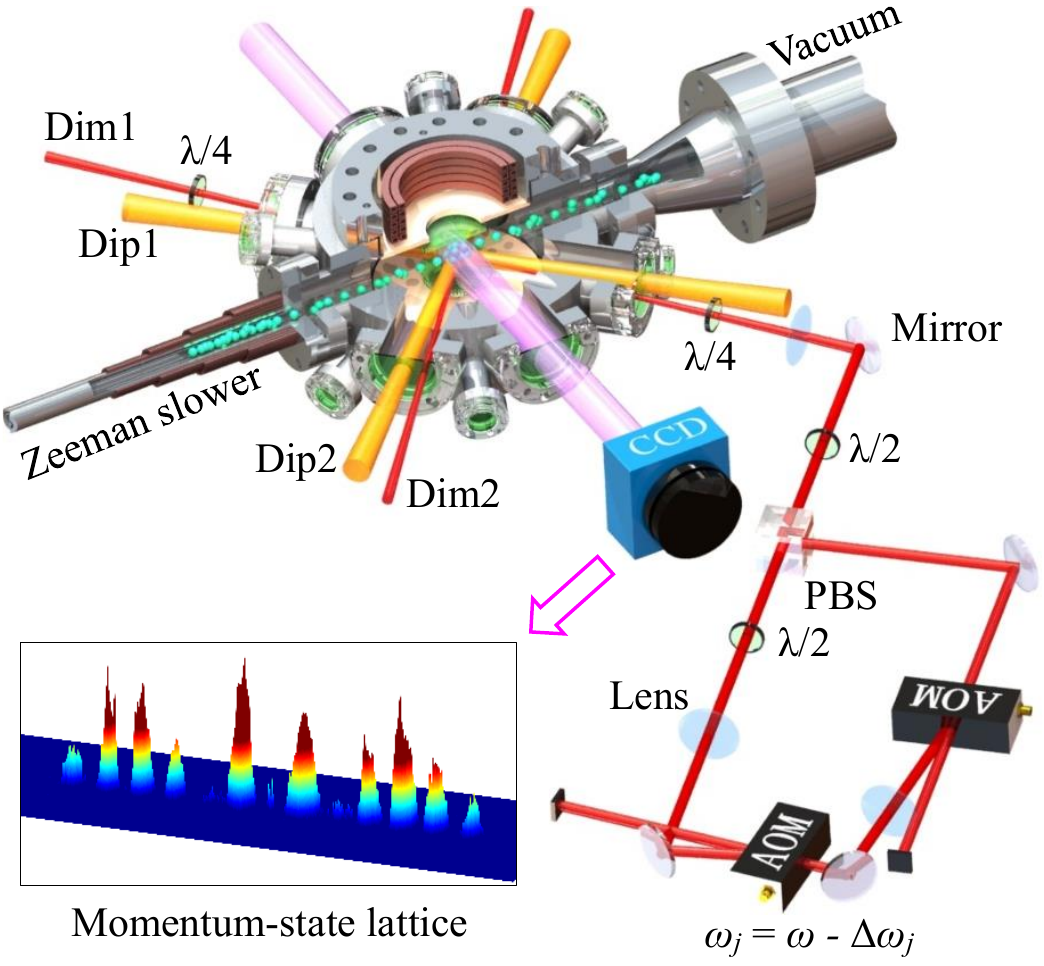}
	\caption{\label{FigS1} The experimental setup for implementing the momentum-state lattice in a quasi-1D $^{133}$Cs BEC. We have used the following abbreviations: (1) Dip: dipole trap laser, (2) Dim: dimple trap laser, (3) PBS: polarization beam splitter, (4) AOM: acousto-optic modulator, (5) $\lambda/4$: quarter-wave plate, (6) $\lambda/2$: half-wave plate. The inset illustrates the distribution of noninteracting atoms in a momentum-state lattice with tunneling amplitude $J/\hbar =2\pi\times 500$ Hz after an evolution time of $0.96$ ms. The absorption image is taken after 18 ms time of flight.}
\end{figure}

Here we provide details on the experimental preparation of the quasi-1D interacting gas of ultracold $^{133}$Cs atoms and the realization of the quasi-periodic momentum-state lattice. An overview of the experimental setup is shown in Fig.~\ref{FigS1}.

By using three-dimensional degenerated Raman sideband cooling, $^{133}$Cs atoms are cooled to a low temperature of $\sim$ 1 $\mu$k, where the atomic sample is polarized in the $|F = 3$, $m_{F} = 3\rangle$ state. The cooled $^{133}$Cs atoms are then loaded into a magnetically levitated dipole trap, which is composed of a magnetic gradient $\partial B / \partial z = 31.3$ G/cm and two orthogonal 1064 nm dipole trap lasers (Dip1 and Dip2) focused to 1/$e^{2}$ radii of 300 $\mu$m~\cite{Wang2021,Weber2003,Kraemer2004,Hung2008}. A crossed dimple trap, formed by another two crossed 1064 nm dimple trap lasers (Dim1 and Dim2) with the 1/$e^{2}$ radii of 58 $\mu$m, is overlapped with the dipole trap to improve the atomic density by a factor of $\sim 30$. A hybrid evaporative cooling is implemented by reducing the magnetic field gradient to $\partial B / \partial z = 0$ and lowering the powers of three lasers (Dip1, Dip2 and Dim2), while the power of the other dimple trap laser (Dim1) is increased to provide a strong radial confinement. Finally, a BEC of $\sim 4\times10^{4}$ Cs atoms is prepared in a quasi-1D optical trap that is formed by two laser beams (Dim1 and Dip2) with a crossing angle of $\sim 80^{\textrm{o}}$. The $z$ axis is defined along the propagation direction of the remaining dimple trap laser (Dim1), the trap frequencies are ($\omega_{x}$, $\omega_y$, $\omega_z$) = 2$\pi$ $\times$ (125, 96, 13)~Hz.

To create a 1D momentum-state lattice, we employ the techniques in Ref.~\cite{Meier2016}. A fraction of the remaining dimple trap laser (Dim1) is retro-reflected to construct a pair of counter-propagating lasers by using a polarization beam splitter (PBS) to combine with the incident laser. The retro-reflected laser is passed through two acousto-optic modulators (AOMs) with diffraction orders ($\pm 1$), where the one AOM is driven with a single-frequency rf signal but the other is driven with the 20-frequency rf signals created by an arbitrary waveform generator. Thus, in compared to the incident laser with the single frequency $\omega$, the reflected laser contains multi-frequency spectral components $\omega_j = \omega -\Delta \omega_j$ with $j = -10,...,9$ [see Fig.~1(b) in main text]. By engineering the detuning of multi-frequency components with programmable rf signals, the counter-propagating laser pair drives a series of two-photon Bragg resonance transitions, which couple the discrete momentum states with a momentum difference $2\hbar k$. Because the Bragg lasers are far detuned from the atomic transition between the ground and excited states ($|g\rangle$ and $|e\rangle$), the rotating wave approximation is used to adiabatically eliminate $|e\rangle$. As a result, the Bragg laser pair induces a time-dependent optical potential
\begin{equation}
V_\textrm{Brag}(z, t)=2J\sum_{j}\cos(2kz-\Delta \omega_j t+\theta_j), \label{eq:Vbragg}
\end{equation}
where $J$ is obtained from the two-site Rabi oscillation and $\theta_j$ is the relative phase.

To implement the incommensurate modulations as in the AA model, we design $\hbar \Delta \omega_j=4(2j+1)E_R-(\epsilon_{j+1}-\epsilon_j)$, where $E_R = \hbar^2 k^2/2m= h \times 5.3$ kHz is the recoil energy and $\epsilon_{j} = \Delta \cos(2\pi \beta j + \phi)$ with $\beta=(\sqrt{5}-1)/2$~\cite{An2017}. Thus, by tuning the strength $J$, detuning $\Delta \omega_j$ and phase $\theta_j$ involved in every Bragg transition, we can directly control over the amplitude and phase of the coupling between the adjacent lattice sites, as well as the site energy in the momentum lattice. In all the measurements, the tunneling amplitude $J$ has been regularly calibrated by using two-site Rabi oscillations, and the errors are ensured to be less $5\%$.  \\

\subsection{Measurement of the momentum distribution}

Here we provide detailed description of the measurement for the momentum distribution of atoms in the main text.

In the measurements of atomic mean momentum width and participation ratio for various disorder and interaction, we switch off all laser fields and quickly tune the external magnetic field to $17$~G with zero-crossing scattering length. After $18$~ms time of flight, we detect the populations of atoms in different momentum states via standard absorption imaging technique as shown in the inset of Fig.~\ref{FigS1}. The imaging laser beam passes through the atomic cloud at an intersecting angle of $35^{\textrm{o}}$ with respect to the direction of atomic momentum transfer. The momentum distribution of atoms can then be identified with different momenta.\\

\begin{figure}
	\includegraphics[width=0.97\columnwidth]{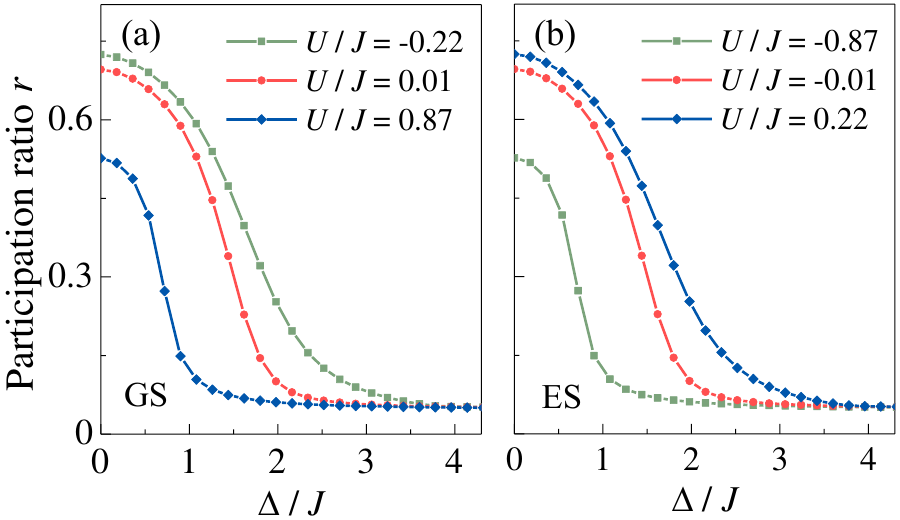}
	\caption{\label{FigS2} Theoretical participation ratio $r$ as a function of disorder $\Delta/J$ for various interaction $U/J$. The value of $r$ is obtained based on the nonlinear AA model under various disorder $\Delta/J$ and interaction $U/J$. Results are shown for (a) the ground state (GS) and (b) the highest excited state (ES), respectively. We take the same system parameters as in Figs. 4(a) and (b) in the main text. }
\end{figure}

\subsection{Mean-field theory for the momentum-state lattice system}

Following Refs.~\cite{Chen,An2018,An2021b}, here we briefly recapitulate the Gross-Pitaevskii (GP) theory for the momentum-state lattice.

For a BEC in the combined presence of a harmonic trap potential $V_\textrm{trap}(x,y,z)$ and the optical potential of the Bragg lasers $V_\textrm{Brag}(z, t)$, it can be described in the mean-field theory by a 3D wavefunction $ \Psi({\bf r},t)$ which satisfies the GP equation
\begin{equation}
i\hbar\frac{\partial \Psi}{\partial t}=\Big(-\frac{\hbar^2}{2m}\nabla^2+V_\textrm{trap}+V_\textrm{Brag}+\frac{4
\pi\hbar^2 a}{m}N|\Psi|^2\Big)\Psi, \label{eq:GP0}
\end{equation}
with the renormalization condition $\int d{\bf r}|\Psi({\bf r},t)|^2=1$. Here, $N$ is the total atom number. Based on Eq.~(\ref{eq:GP0}), we calculate mean momentum width $\langle d\rangle$ by taking the corresponding experimental parameters. The theoretical results shown in Figs.~3(b)-(c) in the main text, which are scaled with a factor to account for the effects of decoherence, agree qualitatively well with the experimental data.

Next, as the lattice is periodic, one assumes the ansatz $\Psi({\bf r},t)=\sum_{j}\psi_j({\bf r},t)e^{i(2jkz-4j^2E_Rt/\hbar)}$. Here the mode function $\psi_j({\bf r},t)$ is narrowly localized at $p_j=2j\hbar k$ in the momentum space. Consider first the shallow harmonic trap is ignored. The equation for $\psi_j$ can be readily derived in the standard variational approach by minimizing the BEC energy functional in the presence of $V_\textrm{Brag}(z, t)$, yielding
\begin{eqnarray}
i\hbar\frac{\partial\psi_j}{\partial t}&=&\Big(-\frac{\hbar^2\nabla^2}{2m}-\frac{2j\hbar k}{m}i\hbar\partial_z\Big)\psi_j\nonumber\\
&+&J\psi_{j+1}+J\psi_{j-1}+\Delta \cos(2\pi \beta j+\phi)\psi_{j}\nonumber\\
&+&\frac{4
\pi\hbar^2 a}{m}N(2\sum_{l\neq j}|\psi_{l}|^2+|\psi_{j}|^2)\psi_{j}.
\end{eqnarray}
By assuming a $j$-independent homogenous density distribution so that $\psi_j=\tilde{\psi}_j/\sqrt{V_0}$~\cite{Chen,An2018,An2021b}, one obtains
\begin{eqnarray}
i\hbar\frac{\partial\tilde{\psi}_j}{\partial t}&=&J\tilde{\psi}_{j+1}+J\tilde{\psi}_{j-1}+\Delta \cos(2\pi \beta j+\phi)\tilde{\psi}_{j}\nonumber\\
&+&U(2-|\tilde{\psi}_{j}|^2)\tilde{\psi}_{j}, \label{eq:nonlinear AA_GP}
\end{eqnarray}
with the normalization condition $\sum_j|\tilde{\psi}_{j}|^2=1$. In Eq.~(\ref{eq:nonlinear AA_GP}), $U=(4\pi\hbar^2a/m)\bar{\rho}$ is the homogeneous mean-field energy, where $\bar{\rho}=N/V_0$ denotes the average density. Then, taking into account of harmonic trap, the local density approximation is assumed and an effective homogeneous density $\bar{\rho}$ is evaluated using the Thomas Fermi radius. Equation~(\ref{eq:nonlinear AA_GP}) is formally the same as Eq.~(1) in the main text. In the experimental localization measurement in Fig.~4 of main text, we have used $V_0=\int_{|\Psi({\bf r},t=0)|\ge 0.005}d{\bf r}$ and thus $\bar{\rho}=2\times 10^{13} \textrm{cm}^{-3}$ to calculate $U$.

We comment that the above GP description including the underlying assumptions has proven to be a good description of the momentum-lattice systems in previous atomic experiments where $a>0$~\cite{An2018a,An2021a,An2017,An2018,An2021b}. Here, our result shows that, for $a<0$, Eq.~(\ref{eq:nonlinear AA_GP}) remains a good description on the time scale relevant for our experiments.  \\

\begin{figure}
	\includegraphics[width=0.97\columnwidth]
	{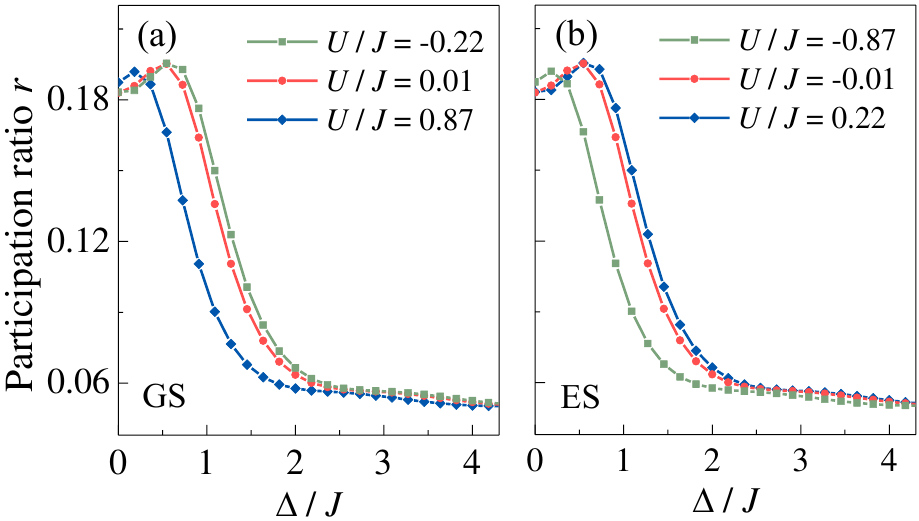}
	\caption{\label{FigS3} Non-adiabatic effects due to the experimental ramp with a finite velocity. Participation ratios $r$ of (a) the ground state (GS) and (b) the excited state (ES) are calculated based on the numerical solutions of Eq.~(\ref{eq:nAAt}) that simulate the ramp process. Specifically, we consider $J(t)=2\pi \times \hbar v t$ during the time interval $t\in [0,1]$ ms, with a ramp velocity $\upsilon= 275~\textrm{Hz}/\mathrm{ms}$. Other system parameters are the same as Figs.~4(a) and (b) in the main text.}
\end{figure}

\subsection{Identification of the extended-to-localized transition point}

Here we provide details on how we experimentally identify the extended-to-localized transition point of a state by fitting the data.

After we obtained the measured participation ratio $r$ for GS (or ES) under various interaction $U/J$ and disorder $\Delta/J$, we plot $r$ as a function of disorder $\Delta/J$ at a given interaction $U/J$. We fit the experimental data using the following empirical relation
\begin{eqnarray}
r= A (\Delta/J)^{-\gamma} \Theta [(\Delta-\Delta_c)/J] + B,\label{eq:fit}
\end{eqnarray}
where $A$, $B$, $\Delta_c$ and $\gamma$ are fitting parameters, and $\Theta(x)=1$ for $x<0$ and $\Theta(x)=0$ for $x>0$. We thus read off $\Delta_c$ as the transition point. Repeating similar procedures for various interactions for GS and ES, respectively, we obtain the data for the transition points shown in Fig.~4 of the main text.

\subsection{Non-adiabatic effect in the state preparation}

In this section, we analyze the non-adiabatic effects arising from the finite ramp velocity in the experiment.

We first note that the participation ratio $r$ calculated from the nonlinear AA model, as shown in Fig.~\ref{FigS2}, corresponds to the ideally adiabatic situations. We see that the experimental value of $r$ is notably smaller than the ideal value. To understand the difference, we take into account of the finite velocity of the ramp in the experiment, and numerically solve the following equation
\begin{eqnarray}
\label{eq:nAAt}
i\hbar\dot{\varphi}_{j}&=&J(t)(\varphi _{j+1}+\varphi _{j-1})+\Delta \cos(2\pi\beta j+\phi)\varphi _{j}\nonumber\\
&-&U|\varphi _{j}|^{2}\varphi _{j}.
\end{eqnarray}
Here, $J(t)=2\pi \times \hbar v t$ for $t\in [0,1]$ ms with $\upsilon= 275 ~\textrm{Hz}/\mathrm{ms}$ simulates the experimental ramp process. In the following, we describe the calculations for GS as an example. For a given value of disorder $\Delta$ and interaction $U$, we take the initial state as the GS of the nonlinear AA model with $J=0$. Next, we numerically solve Eq.~(\ref{eq:nAAt}) to obtain ${\varphi}_{j}(t_f)$ at $t_f=1$~ms, and calculate its participation ratio $r$. The resulting $r$ is plotted as a function of $\Delta/J$ for three different $U/J$ in Fig.~\ref{FigS3}(a). Similarly, we obtain the results for ES in Fig.~\ref{FigS3}(b). Compared to the ideal case in Fig.~\ref{FigS2}, we see that, indeed, the non-adiabatic effect provides a main contribution to the decrease in the participation ratio $r$, particularly in the regime of weak disorder. We emphasize, however, that such imperfection does not qualitatively change the key result, i.e., interaction enhances or suppresses the localization of an eigenstate depending on its energy. \\

\begin{figure}
	\includegraphics[width=1\columnwidth]
	{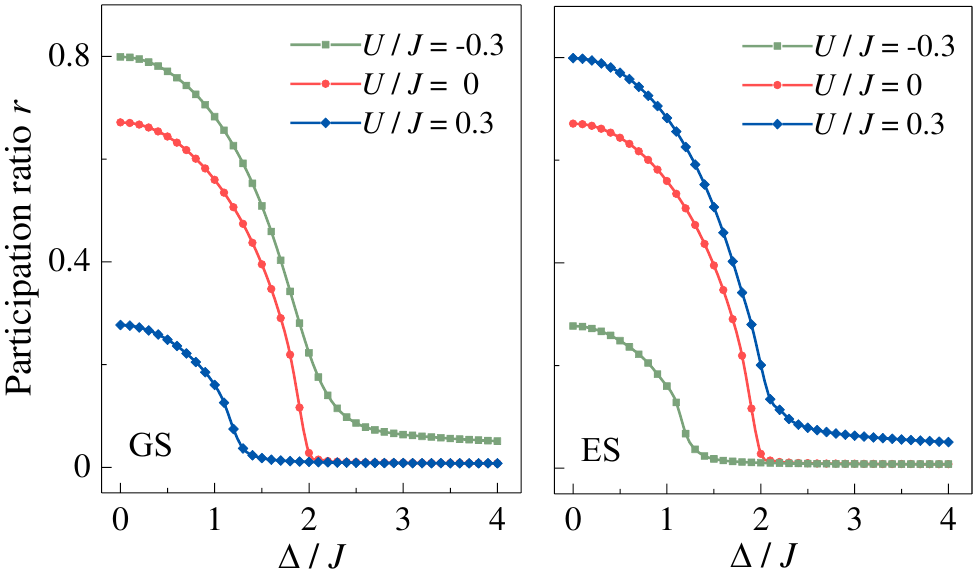}
	\caption{\label{FigS4} Participation ratios $r$ of GS and ES, respectively, calculated for a nonlinear AA model with $L$ = 144 lattice sites. Results are shown as a function of disorder $\Delta/J$ for various interaction $U/J$.}
\end{figure}

\subsection{Finite-size effect}

Here, we theoretically study a nonlinear AA model with a size much larger than the experiment and show that the finite-size effect does not qualitatively change the physical picture, i.e., weak interaction suppresses or enhances the localization of an eigenstate depending on the energy, giving rise to ME.

Figure~\ref{FigS4} presents the calculated participation ratio $r$ of GS and ES for a nonlinear AA model with $L=144$ sites under various disorder and interaction. Compared to Figs.~2(a)-(b) in the main text, we see that the size effect quantitatively changes the values of participation ratios and hence transition points. However, it does not qualitatively alter the key features, i.e., the localization of GS (ES) is enhanced (suppressed) for $U>0$, whereas the opposite occurs for $U<0$, which is regardless of the system size.

~\\